\documentclass[aps,prx,twocolumn,reprint,floatfix, superscriptaddress]{revtex4-2}

\usepackage{leftindex}
\usepackage[normalem]{ulem}
\usepackage[utf8]{inputenc}
\usepackage{graphicx,amsmath,amssymb,braket}
\usepackage{color}

\definecolor{darkgreen}{rgb}{0.0, 0.4, 0.26}

\usepackage[colorlinks=true,citecolor=myblue,linkcolor=myred]{hyperref}

\definecolor{mygrey}{gray}{0.35}
\definecolor{myblue}{rgb}{0.2,0.2,0.8}
\definecolor{myzard}{cmyk}{0,0,0.05,0}
\definecolor{mywhite}{rgb}{1,1,1}
\definecolor{mywhite}{rgb}{1,1,1}
\definecolor{myred}{rgb}{1,0.,0.3}

\def\be{\begin{equation}}
\def\ee{\end{equation}}
\def\ba{\begin{align}}
\def\enda{\end{align}}
\def\bi{\begin{itemize}}
\def\ei{\end{itemize}}

\def\beq{\begin{equation}}
\def\beq{\begin{equation}}
\def\eeq{\end{equation}}

\begin{document}

\title{Engineering giant transmon molecules as mediators of conditional two-photon gates}

\author{Tom\'as Levy-Yeyati}
\email{tomas.levy@iff.csic.es}
\affiliation{Institute of Fundamental Physics IFF-CSIC, Calle Serrano 113b, 28006 Madrid, Spain.}
\author{Tom\'as Ramos}
\affiliation{Institute of Fundamental Physics IFF-CSIC, Calle Serrano 113b, 28006 Madrid, Spain.}
%\author{Juan José García-Ripoll}
%\affiliation{Institute of Fundamental Physics IFF-CSIC, Calle Serrano 113b, 28006 Madrid, Spain.}
\author{Alejandro Gonz\'alez-Tudela}
\email{a.gonzalez-tudela@csic.es}
\affiliation{Institute of Fundamental Physics IFF-CSIC, Calle Serrano 113b, 28006 Madrid, Spain.}

\begin{abstract}
Artificial atoms non-locally coupled to waveguides -- the so-called giant atoms-- offer new opportunities for the control of light and matter. In this work, we show how to use an array of non-locally coupled transmon ``molecules" to engineer a passive photonic controlled gate for waveguide photons.  In particular, we show that a conditional elastic phase shift between counter-propagating photons arises from the interplay between direction-dependent couplings, engineered through an interplay of non-local interactions and molecular binding strength; and the nonlinearity of the transmon array. We analyze the conditions under which a maximal $\pi$-phase shift --and hence a CZ gate-- is obtained, and characterize the gate fidelity as a function of key experimental parameters, including finite transmon nonlinearities, emitter spectral inhomogeneities, and limited cooperativity. Our work opens the use of giant atoms as key elements of microwave photonic quantum computing devices.
\end{abstract}

\maketitle

Experimental advances in waveguide QED setups in the microwave are enabling the coupling of artificial atoms, like transmons, to distant apart positions in the waveguide~\cite{gustafsson14a,Kannan2020,kannan2023,joshi2023,Almanakly2024}. Such setups, generally labeled as giant atoms~\cite{FriskKockum2021}, lead to a new light-matter coupling regime far from the local dipole approximation where most quantum optical systems work, thus opening new avenues for the control of light and matter degrees of freedom. For the matter degrees of freedom, the interference between the emission in the multiple coupling points have already been shown to lead to non-trivial frequency-dependent relaxation rates~\cite{Gonzalez-Tudela2019b,kockum14a} or decoherence-free and non-trivial emitter interactions~\cite{kockum18a,Soro2023,Leonforte2024,Wang2021a,Vega2021a,Ingelsten2024,Chen2025}, with applications in superconducting many-body quantum simulation~\cite{Chen2025}. From the photonic point of view, on the other hand, these non-local couplings have been proposed to obtain directional-dependent emission~\cite{ramos16a,Soro2022,Vega2023TopologicalQED,Du2025}, as recently implemented in the lab~\cite{kannan2023,joshi2023}, with applications in photonic entanglement generation and distribution~\cite{Almanakly2024,Du2025}. However, all these photonic studies lie in the single photon or linear regime, limiting the potential impact of the non-local coupling to control many-body quantum states of light. Motivated by this limitation, recent works are calculating the modification of two-photon correlations via their scattering with giant atoms~\cite{Cheng2023a,Gu2024b,xue2025twophotonscatteringwaveguidegiant}. However, a more clear application of these non-local light-matter couplings into the photonic many-body regime is still lacking.

\begin{figure}[tb]
    \centering
    \includegraphics[width=0.9\linewidth]{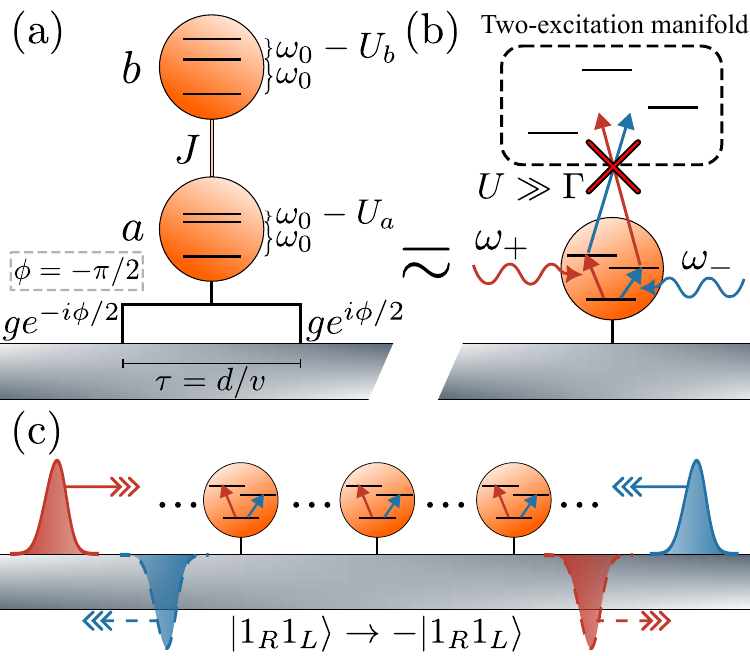}
    \caption{(a) Schematic representation of the building block of the array: a giant transmon molecule formed by two transmons capacitively coupled. One transmon is non-locally coupled to the waveguide at two points with the same strength but opposite complex phase $\phi=-\pi/2$. (b) Single and two excitation Hilbert space of the transmon molecules. At the single photon level, the molecule behaves as a three-level system in the V-configuration, where each level couples only to right- or left-propagating photons if the non-local couplings are properly designed. The two excitation subspace is formed by three states, but for sufficient anharmonicity its excitation is supressed. (c) Setup for the conditional gate: the photonic qubits are encoded in left and right moving photons, which interact with an array of transmon molecules.}
    \label{fig:1}
\end{figure}

In this work, we provide such application by showing how an array of non-locally coupled transmons molecules~\cite{Aamir2022}, see Fig.~\ref{fig:1}(a), to a waveguide can be used to obtain an elastic, conditional $\pi$-shift between counter-propagating photons. This is achieved by a non-trivial combination of three phenomena, that are, i) a state-dependent directional coupling that emerges at a particular choice of the non-local light-matter ($g e^{\pm i\phi/2}$) and transmon couplings ($J$)~\cite{Du2025}; ii) the inhibition of two-excitation molecular states due to the non-linearity of the transmons and the molecular selection rules, see Fig.~\ref{fig:1}(b); iii) the filtering of inelastic scattering processes due to the non-linear polariton dispersion emerging when using an array of multiple molecules~\cite{brod2016,Schrinski2022,LevyYeyati2025}. Here, we first demonstrate the ideal conditions under which these three elements lead to a conditional $\pi$-shift that can be used to engineer a deterministic CZ gate between photonic waveguide qubits. Then, we characterize the gate fidelity in terms of the relevant experimental parameters such as transmon non-linearity (demonstrating that one of the transmons of the molecule can be replaced by a resonator), non-guided decay, and spectral transmon inhomogeneities, showing it is within the reach of state-of-the-art platforms.

\emph{Setup.-} Our proposal consists of $N$ transmon molecules, where one of the elements of the dimers is non-locally coupled to the microwave waveguide, see Fig.~\ref{fig:1}(a). Each molecule is composed of two transmons, labeled as $a$ and $b$, both with transition frequency $\omega_0$ and non-linearity $U_a$ and $U_b$, respectively. The pair of transmons are capacitively coupled with strength $J$, that we assume to be $J\ll \omega_0$, such that we can neglect the counter-rotating terms of the coupling Hamiltonian. Under this limit, and setting $\hbar\equiv 1$ for the rest of the manuscript, the Hamiltonian of each molecular element is composed by a linear part, $\hat H_M=\omega_0\  \left(\hat a^\dagger \hat a+\hat b^\dagger \hat b\right) +J\left( \hat a^\dagger \hat b + \hat b^\dagger \hat a\right)$, and a non-linear term: $\hat V=-\frac{U_a}{2}\hat a^\dagger \hat a\left( \hat a^\dagger \hat a -1 \right)-\frac{U_b}{2}\hat b^\dagger \hat b\left( \hat b^\dagger \hat b -1 \right)$, being $\hat a^\dagger$ ($\hat b^\dagger$) the creation operator of an excitation in $a$ ($b$) transmon. As shown in Fig.~\ref{fig:1}(a), the coupling to the waveguide photons, that we assume to be linearly dispersive $\omega_k=v|k|$ and described by creation [destruction] operators $\hat c_k^\dagger$ [$\hat c_k$], $\hat H_B=\int dk\ \omega_k\ \hat c^\dagger_k\hat c_k$; has to be done only through one of the transmons, e.g., the $a$ one. In that case, the light-matter interaction Hamiltonian is given by:
\begin{equation}\label{eq.Hi}
    \hat H_I=\int dk\ g(k)\ \hat a^\dagger \hat c_k + \text{H.c.}\,,
\end{equation}
where we again neglect the counter rotating terms, and consider a $k$-dependent coupling, $g(k)$, that depends on the non-local couplings of the $a$ transmon to the waveguide. For example, assuming that $a$ couples to two waveguide positions separated a distance $d$ with coupling strength $ge^{\pm i\phi/2}$, as depicted in Fig.~\ref{fig:1}(a), $g(k)$ reads~\cite{ramos16a}:
\begin{equation}\label{eq:coupling}
    g(k)=\sqrt{\frac{2}{\pi}}g\cos\left( \frac{kd+\phi}{2}\right)\,,
\end{equation}
where we do not write explicitly a global phase $e^{ikx_n}$ encoding the central position of the dimer $x_n$. In what follows, we will analyze both the single and two-photon scattering properties of such elements, and show how such simple configuration can be used to engineer a conditional $\pi$-phase between counter propagating photons.

\begin{figure}[t!]
    \centering
    \includegraphics[width=0.95\linewidth]{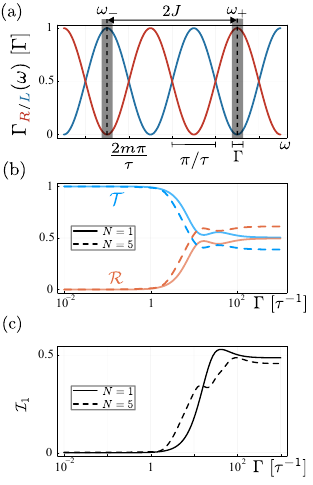}
    \caption{(a) Effective decay rate $\Gamma_{R/L}$ to right- (red) and left- (blue) propagating modes as a function of the frequency as defined in Eq.~\eqref{eq:freqdepchir}. Gray shaded areas correspond to a situation where the single-photon resonances of the molecules, $\omega_{\pm}$ achieve a state-dependent chirality. (b) Transmittance $\mathcal{T}$ (blue), reflectance $\mathcal{R}$ (orange) and (c) infidelity $\mathcal{I}_1$ of the overlap with an ideal chirally transmitted pulse, as defined in Eq.~\eqref{Eq:I1} of a single-photon right-(left-)propagating pulse resonant with $\omega_+$($\omega_-$), as a function of $\Gamma$. The pulse bandwidth is set to $\sigma=0.1\ \Gamma$ to resolve each resonance. We set $\omega_0\tau=21\pi$, $J\tau=\frac{3}{2}\pi$, and the distance between the emitters to be $D=5v\tau$. For those and for other distances and suitable parameters, we see that the chiral regime is reached when $\Gamma\tau\lesssim 0.1$, $\sigma\lesssim 0.1\ \Gamma$.}
    \label{fig:2}
\end{figure}

\emph{Single photon regime.-} First, notice that, if we neglect the coupling to the $b$ element, i.e., $J=0$, such non-local coupling configuration with $\phi=-\pi/2$ is what has been recently implemented experimentally to achieve a directional emission with a single transmon~\cite{kannan2023,joshi2023,Almanakly2024}. This can be understood by calculating the effective decay rates into the left and right-moving photons as a function of the frequency, $\Gamma_{R/L}(\omega)=\frac{\pi}{v}|g(\pm\omega/v)|^2$, which reads:
\begin{equation}\label{eq:freqdepchir}
       \Gamma_{R/L}(\omega)=\frac{\Gamma}{2}\left[1\pm\sin(\omega\tau)\right]\,,
\end{equation}
with $+$ for $R$ and $-$ for $L$, being $\tau=d/v$ the time it takes to the photons to travel between the two coupling points and $\Gamma=2|g|^2/v$. In the uncoupled case, this means that for specific conditions of the frequency $\omega_0$ and distance $d$, the coupling of the $a$ transmon can be made perfectly chiral, see Fig.~\ref{fig:2}(a).

In the molecular case, i.e., when $J\neq 0$, the situation can be richer~\cite{Du2025}. The reason is that when $J\neq 0$, there are two different single photon transitions at energies $\omega_{\pm}=\omega_0\pm J$. Thus, choosing the system parameters such that:
\begin{equation}\label{eq:resonances}       
\omega_\pm\tau=\left(2m_{\pm}\pm\frac{1}{2}\right)\pi\,,
\end{equation}
with $m_\pm\in \mathbb{Z}$, and $\Gamma\tau\ll1$, which together implies $J\gg\Gamma$, each resonance is well resolved in a frequency region with opposite chirality, and then the system features a state-dependent chirality, see Fig.~\ref{fig:2}(a). This means that, at the single-photon level, the molecule behaves effectively as a $V$-level system with each transition coupled only to  left or right moving photons, respectively, see Fig.~\ref{fig:1}(b). In the ideal case, the reflection from both sides of the molecule is suppressed due to the chirality, and thus a the single photon scattering matrix after passing through the $N$ molecules reads: $\hat S_N\ket{\omega}_\mu\approx t^N_\mu(\omega)\ket{\omega}_\mu$, with $\mu=\{R,L\}$ and $\ket{\omega}_{R(L)}$ represents a single right-(left-)propagating photon of frequency $\omega$, and:
\begin{equation}\label{eq:transmission}
    t_{R/L}(\omega)\approx\frac{\omega-\omega_{\pm}-i\frac{\Gamma}{2}}{\omega-\omega_{\pm}+i\frac{\Gamma}{2}}\,,
\end{equation}
with $+$ for $R$ and $-$ for $L$. For a single incoming right-(left-)propagating foton of frequency $\omega_+$($\omega_-$), with a small bandwidth wavepacket, $\sigma\ll\Gamma$, this imparts a phase-shift $\pi$ and a delay $T=4/\Gamma$ for every molecule the wavepacket encounters \cite{Mahmoodian2020}. Deviating from the ideal case, there will be reflection and deformation in the transmission of the wavepacket, which we also compute with an exact transfer matrix formalism~\cite{Niemet2020} using the exact transmission and reflection coefficients obtained in Appendix~\ref{app:1}~\cite{Piasotski2021,Gu2024b}. In particular, we calculate the single-photon scattering response for $N$ transmon molecules separated at distances $D$ between them, 
considering possible deviations from the ideal setup. The results of this analysis are summarized in Figs.~\ref{fig:2}(b-c). First, in Fig.~\ref{fig:2}(b), we plot both the global transmission ($\mathcal{T}$) and reflection ($\mathcal{R}$) of the molecular array in a state-dependent chirality condition as a function $\Gamma$. As expected, the ideal behaviour $\mathcal{T}\approx 1$ is achieved when $\Gamma\tau\lesssim 0.1$ and $J\gtrsim 10\ \Gamma$, that is, when the peaks are well resolved, and the chiral condition is satisfied for all the frequencies of the wavepacket involved. 

For the gate proposal, it is not only important that the single photons coming from the left and right do not reflect, but also that they are not deformed and acquire a $\pi$ phase-shift for each molecule of the array. For this reason, we also plot in Fig.~\ref{fig:2}(c) the following single photon infidelity:
\begin{equation}\label{Eq:I1}
    \mathcal{I}_1=\frac{1}{2}\left| 1- \text{Re}\left(_\mu \braket{\psi_\text{id,out} |\psi_\text{out}}_\mu\right) \right |\,,
\end{equation}
computed from the overlap of the exact output with the ideal single-photon chiral transfer, i.e.
\begin{equation}
    \ket{\psi_{\text{id,out}}}_\mu=\frac{(-1)^N}{\sqrt{\sigma\sqrt{2\pi}}}\int d\omega\ e^{-\frac{(\omega-\omega_{\pm})^2}{4\sigma^2}}e^{iNT(\omega-\omega_\pm)} \ket{\omega}_\mu\ ,
\end{equation}
with $\mu=R [L]$ for $\omega_{+[-]}$, respectively. This infidelity is $0$ if the photons acquire the ideal phases ($e^{i\pi}$) and delays ($T=4/\Gamma$) for each molecule that it encounters, $0.5$ if the photon is completely reflected, and $1$ if is transmitted and delayed but does not acquire the $\pi$-phases. Again, we numerically confirm that the array behaves ideally for the gate when the resonances meet Eq.~\eqref{eq:resonances} and $\Gamma\tau\lesssim 0.1$.

\begin{figure}[tb]
    \centering
    \includegraphics[width=0.95\linewidth]{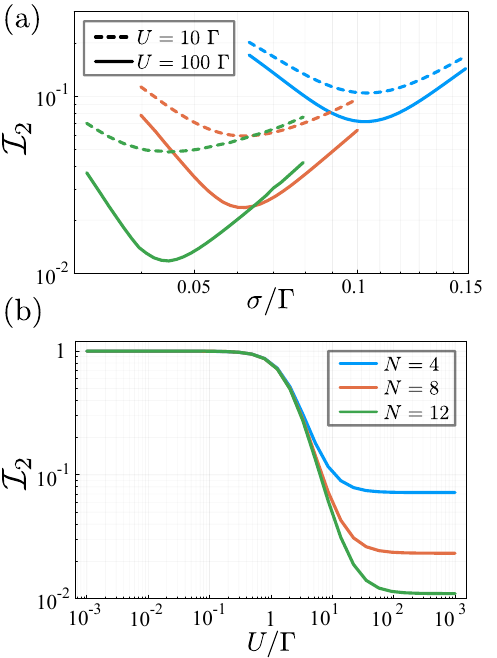}
    \caption{(a-b) Infidelity of the two-photon process $\mathcal{I}_2$ as defined in Eq.~\eqref{eq:Inf} as a function of the input bandwidth $\sigma/\Gamma$ and anharmonicity parameter $U/\Gamma$, respectively. In (a), the different colors [lines] correspond to different numbers of molecules, $N=\{4,8,12\}$ (blue, orange,  green) [anharmonicities $U/\Gamma=\{10,100\}$ (dashed, solid)]. In (b), the colors correspond to different molecule numbers $N=\{4,8,12\}$, choosing the optimal band-width $\sigma/\Gamma=\{0.1, 0.06, 0.045\}$ (blue, orange, green).}
    \label{fig:3}
\end{figure}
\textit{Two-photon~regime.-} In the state-dependent chiral regime previously analyzed, one can envision a translation of the photon gate between the counter-propagating photons proposed for $V$-level systems in Ref.~\cite{Schrinski2022}. There, the saturation of their emitters makes the photons acquire a $\pi$-shift less when the two photons meet in the array. However, our transmon molecules have a two-excitation subspace with three additional levels which does not guarantee translating the aforementioned proposal is possible. To check under which conditions our transmon molecule array can be used to engineer a conditional, elastic phase shift, we now calculate the two-photon scattering process taking into account both this two excitation manifold and the possible inelastic processes. For that, we consider the situation where one right-propagating photon of frequency $\omega_R\sim\omega_+$ and one left-propagating photon of frequency $\omega_L\sim\omega_-$ enter the system, both with a small bandwidth $\sigma\ll\Gamma$, and use as a figure of merit the infidelity of the two photon process, 
\begin{equation}\label{eq:Inf}
    \mathcal{I}_2=\frac{1}{2}\left|1- \text{Re}\left( _{RL}\braket{\psi_{\text{id,out}}|\psi_{\text{out}}}_{RL}\right)\right|\, ,
\end{equation}
defined using the overlap of the two-photon output with the ideal one which undergoes an elastic scattering and a non-linear $\pi$-phase shift $\ket{\psi_{\text{id,out}}}_{RL}=-\ket{\psi_{\text{id,out}}}_R\otimes\ket{\psi_{\text{id,out}}}_L$.

To calculate the resulting output wavepacket, $\ket{\psi_{\text{out}}}$, we use the SLH formalism developed in Ref.~\cite{brod2016a}. There, it was shown that the two-photon counterpropagating scattering through the array in the chiral regime will be given by the sum of each process where the photons interact at site $n$:
\begin{align}\label{eq:totaltwophoton}
    &\bra{\omega'_R\omega'_L}\hat S_N\ket{\omega_R\omega_L}\approx\\
    &\sum_{n=1}^N\left[t(\omega_R')t(\omega_L)\right]^{N-n}\bra{\omega'_R\omega'_L}\hat S\ket{\omega_R\omega_L}\left[t(\omega_R)t(\omega_L')\right]^{n-1}\ ,\nonumber
\end{align}
where $\bra{\omega'_R\omega'_L}\hat S\ket{\omega_R\omega_L}$ is the two-photon, single-site, inelastic scattering amplitude connecting $\omega_R\rightarrow\omega_R'$ and $\omega_L\rightarrow\omega_L'$ \footnote{We allow for the following abuse of notation to simplify the expression: $t(\omega_\mu)\equiv t_\mu(\omega_\mu)$ and $\ket{\omega_R\omega_L}\equiv\ket{\omega_R}_R\otimes\ket{\omega_L}_L$.}, as computed in Appendix \ref{app:2}, assuming only $J\gtrsim10\ \Gamma$. From this calculation, we note that as long as the total non-linearity $U=U_a+U_b$ is large enough, $U\gg\Gamma$, this two-photon scattering matrix can be simplified to:
\begin{align}%\label{eq:twophoton}
    \nonumber
    \bra{\omega'_R\omega'_L}&\hat S\ket{\omega_R\omega_L}\approx t(\omega_R)t(\omega_L)\ \delta(\omega_R-\omega_R')\ \delta(\omega_L-\omega_L')\\
    \nonumber
    &+i\frac{\Gamma^2}{2\pi} \ \frac{\delta(\omega_R+\omega_L-\omega'_R-\omega'_L)}{\left(\omega_R-\omega_++i\frac{\Gamma}{2}\right)\left(\omega_L-\omega_-+i\frac{\Gamma}{2}\right)}\\
    &\times \left[\frac{1}{\left(\omega_R'-\omega_++i\frac{\Gamma}{2}\right)}+\frac{1}{\left(\omega_L'-\omega_-+i\frac{\Gamma}{2}\right)}\right]\,.
\end{align}

Here, there are three noteworthy results. First, the sum of all non-linear processes from Eq.~\eqref{eq:totaltwophoton} leads to a destructive interference of the inelastic scattering component resulting in an elastic, non-linear phase-shift when $N\gg1$~\cite{brod2016a}. This coincides with similar results obtained using a polariton picture~\cite{Schrinski2022}. Second, the non-linear scattering within this regime does not depend on the individual non-linearities, $U_{a/b}$, but only on the total one $U=U_a+U_b$, due to their strong hybridization ($J\gg\Gamma$). This is important because it means one of the transmons of each molecule can be replaced by a linear element (resonator). Third, the scattering does not depend on the ratio $U/J$, even though at the limit $U_a=U_b\gg J\gg \Gamma$ one of the two-excitation states becomes resonant with the two impinging photons with energy $E=2\omega_0$. The reason is that, in this limit, it emerges a molecular selection rule from the probability amplitude of going from the ground state of the two transmons ($\ket{0}$) to the doubly excited state ($\ket{E=2\omega_0}$) when two counterpropagating photons of frequencies $\omega_\pm$ arrive, i.e., $\bra{E=2\omega_0}\hat d^\dagger_{+}\hat d^\dagger_{-}\ket{0}$, where $\hat{d}_{\pm}=(\hat{a}\pm \hat{b})/\sqrt{2}$. In the limit of $U\gg \Gamma$, this two-photon amplitude can be shown to vanish due to a destructive interference effect.

\begin{figure}[t!]
    \centering
    \includegraphics[width=0.95\linewidth]{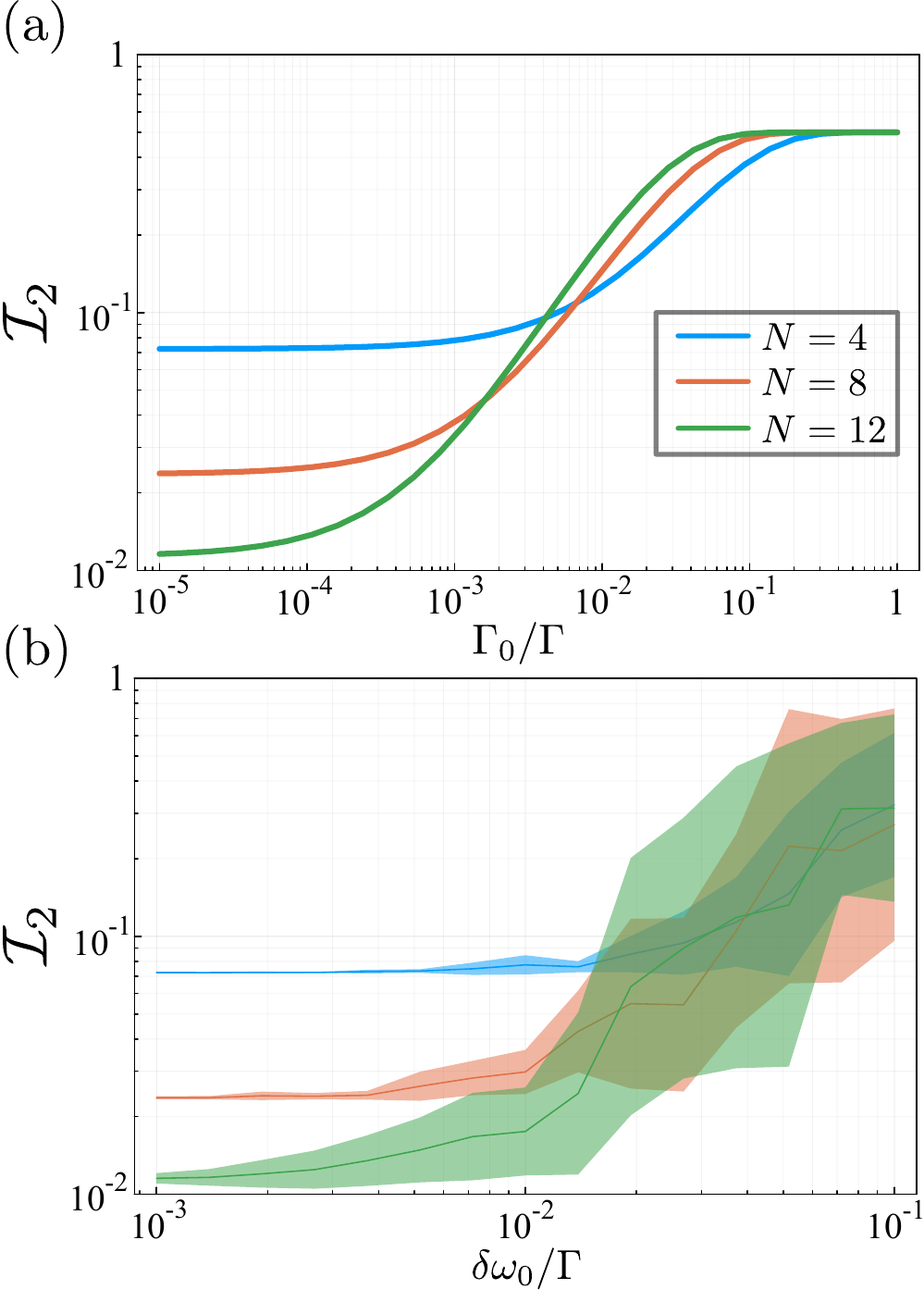}
    \caption{(a) [(b)] Infidelity of the two-photon process as defined in Eq.~\eqref{eq:Inf} as a function of the loss into unwanted channels, $\Gamma_0$ [spectral transmon inhomogeneity $\delta\omega_0$] for $N=\{4,8,12\}$   (blue, orange, green). In (b), we use $15$ random disorder realizations, and plot the geometric mean of the infidelity in solid lines and the multiplicative dispersion as the shaded area. For both (a)-(b), $U=100\ \Gamma$ and, for every $N$, the input bandwidth $\sigma$ is the optimal as in Fig. \ref{fig:3}.}
    \label{fig:4}
\end{figure}

In Fig.~\ref{fig:3}(a), we plot the infidelity of the two-photon process $\mathcal{I}_2$ for different number of transmon molecules $N$ and anharmonicities of the transmons $U$ as a function of the input frequency bandwidth of the photons. There, we observe that there is an ideal frequency bandwidth, which decreases with $N$, independent of the non-linearity. This optimal bandwidth is such that the pulse is short enough in frequency to resolve the resonance, but wide enough to fit spatially inside the array. For the optimal bandwidth, the infidelity decreases with the number of molecules $N$. In Fig.~\ref{fig:3}(b), we calculate the scaling of the infidelity for the optimal bandwidth as a function of the non-linearity $U/\Gamma$ for different number of molecules. When $U\lesssim\Gamma$, the transmons are too harmonic to impart a non-linear phase-shift. For $U\gg\Gamma$, the infidelity reduces, and saturates to the maximal fidelity achievable for a given $N$. 

Finally, in Fig.~\ref{fig:4}(a-b) we also characterize the two-photon infidelity as a function of several experimental imperfections. In particular, in Fig.~\ref{fig:4}(a), we calculate the effect of the decay into other channels different from the waveguide at rate $\Gamma_0$, observing how it increases the infidelity, especially for larger number of emitters. In Fig.~\ref{fig:4}(b), we also compute the $\mathcal{I}_2$ as a function of spectral transmon inhomogeneities. For that, we make the calculation considering a random gaussian disorder over the ideal frequency $\omega_0$ of each transmon, with a standard deviation $\delta\omega_0$. There, we observe how the gate is resilient up to disorder strengths $\delta\omega_0/\Gamma\lesssim~0.01$.

 \emph{Conclusions.-} Summing up, we showed how to engineer a conditional phase gate between counter-propagating photons by using an array of non-locally coupled transmon molecules. The mechanism consists in a combination of a state-dependent directonality appearing at the single photon level, the filtering of inelastic scattering processes due to the periodic nature of the array, and the saturation effects coming from the transmon non-linearities and two photon excitation configuration. By using exact numerical calculations, we characterize the fidelities of the gate at the single and two-photon level under realistic condtions, showing how we can reach fidelities of more than $\gtrsim 90\%$ using only $N=4$ molecules with $U=J=100\Gamma=100$ MHz, and for cooperativities $\Gamma/\Gamma_0\gtrsim 200$, of the order of what has been reported in microwave implementations~\cite{Mirhosseini2019,Sheremet2023WaveguideCorrelations}. Combined with single photon qubit rotations, which can be done with beam splitters in dual rail encodings~\cite{LevyYeyati2025},  our conditional gate offers a universal gate set with which one can generate arbitrary photonic states. Our result places giant atom setups as potential enablers of future microwave photonic quantum devices.

\emph{Acknowledgemts.-} The authors acknowledge support from the CSIC Research Platform on Quantum Technologies PTI-001 and from Spanish projects PID2021127968NB-I00 funded by MICIU/AEI/10.13039/501100 011033/ and by FEDER Una manera de hacer Europa, respectively. A.G.T. also acknowledges support from the QUANTERA project MOLAR with reference PCI2024-153449 andfundedMICIU/AEI/10.13039/ 501100011033 and by the European Union. T.R. further acknowledges support from the Ramón y Cajal program RYC2021-032473-I and the Generaci\'on de Conocimientos project PID2023-146531NA-I00, financed by MCIN/AEI/10.13039/501100011033 and the European Union NextGenerationEU/PRTR. T.L.Y. acknowledges funding from the Ministry of Science, Innovation and Universities of Spain FPU22/02005.

\emph{Data availability.-} The codes used to support the results of this manuscript are openly available \cite{Codes}.

\bibliographystyle{unsrt}
\bibliography{references1,references2}

\appendix

\clearpage

\section{Single-photon scattering}\label{app:1}

In this section, we aim at calculating the general scattering matrix:
\begin{align}
    \label{eqSM:scattering}
    \hat S=\mathbb{I} -2\pi i\ \delta(\omega_f-\omega_i) \ \hat T(\omega_i)\,,
\end{align}
with $\omega_i$ and $\omega_f$ being the asymptotically free input and output energies, and $\hat T(\omega)=\hat H_I+\hat H_I\hat G(\omega)\hat H_I$. Here, $\hat{G}(\omega)$ is the retarded Green's function of the total system, $\hat G(\omega)=(\omega-\hat H+i0^+)^{-1}$, defining $\hat H=\hat H_M+\hat H_B+\hat H_I+\hat V$ as the total light-matter interaction Hamiltonian, and $\hat{H}_I$ only the interaction part. Thus, calculating $\hat S$ boils down to calculating $\hat{G}(\omega)$. For that, we can express $\hat{G}(\omega)$ in terms of the retarded Green's function of the uncoupled system, $\hat G_0(\omega)=(\omega-\hat H_0+i0^+)^{-1}$ with $\hat H_0=\hat H_M+\hat H_B$, using the Lippmann-Schwinger equation:
\begin{align}
    \label{eqSM:Lipmann}
    \hat G(\omega)=\hat G_0(\omega)+\hat G(\omega)\hat H_I\hat G_0(\omega)\,.
\end{align}
valid for single-excitation dynamics, as the nonlinearity $\hat V$ does not matter. 

In the single-excitation subspace and for a single transmon molecule, the scattering matrix for an incoming photon can be shown to be given by 
\begin{align}
    \nonumber
    \bra{k'}\hat S\ket{k}=&\delta(k-k')\\
    &-2\pi i\ \delta(\omega_k-\omega_{k'}) g^*(k')g(k)\bra{a}\hat G(\omega_k)\ket{a}\,,
\end{align}
with $\ket{k}=c_k^\dagger\ket{0}$ and $\ket{\alpha}=\hat{\alpha}^\dagger\ket{0}$ being the states creating one photon with momentum $k$ and one excitation in $\alpha$ transmon, respectively, from the vacuum of excitations $\ket{0}$. Thus, calculating this scattering matrix requires calculating the action of the retarded Green's function on the original transmon states $\bra{a}\hat G(\omega_k)\ket{a}$. For that it is convenient to write $\hat G_0(\omega)$ in the basis of the molecular states, $\ket{\pm}=(\ket{a}\pm\ket{b})/\sqrt{2}$, where it is diagonal, i.e., $\hat{G}_0(\omega)\ket{\pm}=(\omega-\omega_\pm)^{-1}\ket{\pm}$. Using Eq.~\eqref{eqSM:Lipmann}, we can find the retarded Green's function of the complete light-matter system, reading:
\begin{align}\label{eq:Gsinglephoton}
    \nonumber
    \hat G&(\omega)\ket{\pm}=\frac{1}{\omega-\omega_\pm+i\frac{\Gamma}{2}\left( 1+\frac{\omega-\omega_\pm}{\omega-\omega_\mp}\right)}\Bigg[\\
    \nonumber
    &\left(1+i\frac{\Gamma}{2}\frac{1}{\omega-\omega_\mp}\right)\ket{\pm} -i\frac{\Gamma}{2}\frac{1}{\omega-\omega_\mp}\ket{\mp}\\
    &+\frac{1}{\sqrt{2}}\int dk\ \frac{g^*(k)}{\omega-\omega_k+i0^+} \ket{k}\Bigg]\,,
\end{align}
where the self-energy of the emitter, defined by $\Sigma(\omega)=\int dk\ \frac{g^*(k)g(k)}{\omega-\omega_k+i0^+}$, for the conditions chosen in the main text, i.e., $\phi=-\pi/2$, reads: $\Sigma(\omega)=-i2|g|^2/v\equiv -i\Gamma$. This decay rate $\Gamma$ is actually given by the sum of the decay rates into left-, $\Gamma_R(\omega)$, and right-propagating, $\Gamma_L(\omega)$, photons, as defined in the main text in Eq.~\eqref{eq:freqdepchir}.

\begin{figure}
    \centering
    \includegraphics[width=0.99\linewidth]{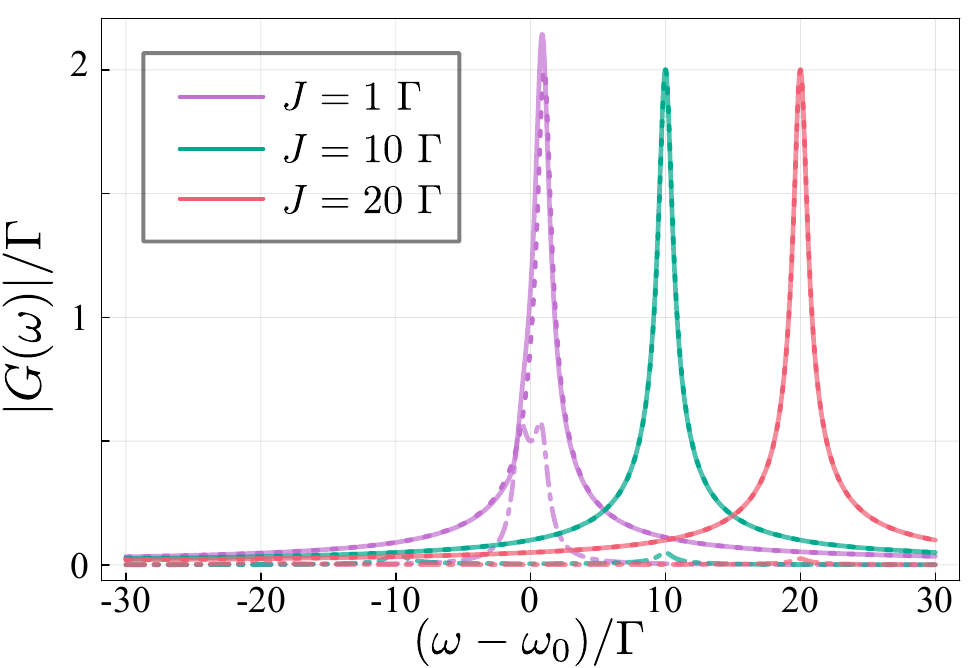}
    \caption{Exact $|\bra{\pm}\hat G(\omega)\ket{+}|$ in solid and dashed-dotted, lines respectively. In dashed lines, we also plot the result of using the approximation described in the main text. The values of $J/\Gamma=\{1,10,20\}$ (purple, green, orange). It is plotted the absolute value of the Green's function.}
    \label{fig:SM1}
\end{figure}

With this we can calculate all the elements in the $\ket{\pm}$ basis of $\bra{a}\hat G(\omega_k)\ket{a}$. Using the states $\ket{\omega}_R\equiv\ket{k>0}$ and $\ket{\omega}_L\equiv\ket{k<0}$ as the single photon basis, we can write the exact scattering matrix of a single transmon molecule $\hat S\ket{\omega}_\mu=t_\mu(\omega)\ket{\omega}_\mu+r(\omega)\ket{\omega}_{\bar\mu}$, where $\mu=\{R,L\}$ and $\bar R=L$, $\bar L=R$. The exact coefficients read:
\begin{equation}
    t_\mu(\omega)=1-i\Gamma_\mu(\omega)\sum_{\nu=\pm}\frac{1}{\omega-\omega_\nu+i\frac{\Gamma}{2}\left(1+\frac{\omega-\omega_\nu}{\omega-\omega_{\bar\nu}}\right)}\ ,
\end{equation}
\begin{equation}
    r(\omega)=-i\sqrt{\Gamma_R(\omega)\Gamma_L(\omega)}\sum_{\nu=\pm}\frac{1}{\omega-\omega_\nu+i\frac{\Gamma}{2}\left(1+\frac{\omega-\omega_\nu}{\omega-\omega_{\bar\nu}}\right)}\ .
\end{equation}
Moreover, in Eq.~\eqref{eq:Gsinglephoton} we can see that for $J\gg\Gamma$, the $\bra{\pm}\hat G(\omega)\ket{+}$ components can simplify.
In Fig.~\ref{fig:SM1}, we calculate the absolute value of these components in solid and dashed-dotted lines, respectively, for different values of $J/\Gamma$. There, we observe that the Green's function controlling the mixing between the $\ket{\pm}$ states is highly suppressed for $J\gtrsim 10\ \Gamma$. Beyond that limit, we find in fact that it is a good approximation to consider: $\bra{\pm}\hat G(\omega)\ket{\pm}\approx(\omega-\omega_\pm+i\frac{\Gamma}{2})^{-1}$, $\bra{\mp}\hat G(\omega)\ket{\pm}\approx 0$ (see dashed lines in Fig.~\ref{fig:SM1}) and:
\begin{equation}\label{eq:aprox1c}
    \bra{k}\hat G(\omega)\ket{\pm}\approx \frac{1}{\sqrt{2}}\frac{1}{\omega-\omega_\pm+i\frac{\Gamma}{2}}\frac{g^*(k)}{\omega-\omega_k+i0^+}\,,    
\end{equation}
to make the calculations. This is the approximation we do to calculate in the two-photon regime.

For the state-dependent chiral regime defined in the main text, we must impose that $ \Gamma_{R}(\omega_{+[-]})=\Gamma\ [0]$ and $\Gamma_{L}(\omega_{+[-]})=0\ [\Gamma]$, obtaining the expected transmission and reflection parameters in the limit $\tau\Gamma\ll1$, where the resonances of bandwidth $\Gamma$ are narrow enough to match the perfectly chiral condition.

\section{Two-photon scattering}\label{app:2}

To solve the two-photon scattering we use the scattering matrix definition of Eq.~\eqref{eqSM:scattering} and find:
\begin{align}\label{eq:B1}
    \nonumber
    &\bra{k'q'}\hat S\ket{kq}=\delta(k-k')\delta(q-q')+\delta(k-q')\delta(q-k')\\
    &\ \ \ \ \ \ -2\pi i\ \delta(E-E')\sum_{p,p'}g^*(\bar p')g(\bar p)\ \bra{a;p'}\hat G(E)\ket{a;p}\,,
\end{align}
where $p=\{k,q\}$, $\bar k=q$, $\bar q=k$ and $E=\omega_k+\omega_q$, $E'=\omega_{k'}+\omega_{q'}$. Moreover, $\ket{\alpha;p}=\hat{\alpha}^\dagger \hat c^\dagger_p\ket{0}$. To calculate it, we also use the Lippmann-Schwinger equation relating the Green's function of the total non-linear system, $\hat H=\hat H_M+\hat H_B+\hat H_I+\hat V$, to the linear Green's function, $\hat H_\ell=\hat{H}_M+\hat H_B+\hat H_I$:
\begin{equation}
    \hat{G}(E)=\hat{G}_\ell(E)+\hat G (E) \hat V \hat G_\ell (E)\ .
\end{equation}
Doing some algebra, we obtain the following equations:
\begin{align}
    \nonumber
    \bra{a;p'}&\hat G(E)\ket{a;p}=\bra{a;p'}\hat G_\ell(E)\ket{a;p}\\
    &-\sum_\alpha U_\alpha\bra{a;p'}\hat G(E)\ket{\alpha\alpha}\bra{\alpha\alpha}\hat G_\ell(E)\ket{a;p}\ ,
\end{align}
\begin{align}
    \nonumber
    &\bra{a;p'}\hat G(E)\ket{\alpha\alpha}=\\
    &\ \ \ \ \ \frac{\bra{a;p'}\hat G_\ell(E)\ket{\alpha\alpha}-U_{\bar \alpha}\frac{\bra{\bar \alpha\bar \alpha}\hat G_\ell(E)\ket{\alpha \alpha}\bra{a;p'}\hat G_\ell(E)\ket{\bar \alpha\bar \alpha}}{1+U_{\bar \alpha}\bra{\bar \alpha\bar \alpha}\hat G_\ell(E)\ket{\bar \alpha\bar \alpha}}}{1+U_\alpha\bra{\alpha\alpha}\hat G_\ell(E)\ket{\alpha\alpha}-U_aU_b\frac{\bra{\bar \alpha\bar \alpha}\hat G_\ell(E)\ket{\alpha\alpha}^2}{1+U_{\bar\alpha}\bra{\bar \alpha\bar \alpha}\hat G_\ell(E)\ket{\bar \alpha\bar \alpha}}} \ ,
\end{align}
where  $\alpha=\{a,b\}$ and $\bar a=b$ and $\bar b=a$. We also define $\ket{\alpha\alpha}=\frac{1}{\sqrt{2}}\hat \alpha^\dagger\hat\alpha^\dagger\ket{0}$ and $\ket{ab}=\hat a^\dagger\hat b^\dagger\ket{0}$.

For the linear system, we can obtain the two-excitation Green's function's elements using the Wick's theorem, which in frequency space, doing a convolution, reads:
\begin{align}
    \nonumber
    \bra{\phi'\psi'}\hat G_\ell(E)\ket{\phi\psi}=&\bra{\phi'}\hat G_\ell(E)\ket{\phi}*\bra{\psi'}\hat G_\ell(E)\ket{\psi}\\
    &+\bra{\psi'}\hat G_\ell(E)\ket{\phi}*\bra{\phi'}\hat G_\ell(E)\ket{\psi}\ ,
\end{align}
where $\phi,\psi,\phi',\psi'$ are any bosonic operators, and taking into account a factor $\sqrt{2}$ whenever $\phi=\psi$ and $\phi'=\psi'$. The convolution is defined: $f(E)*g(E)=\frac{i}{2\pi}\int d\omega\ f(\omega)g(E-\omega)=\frac{i}{2\pi}\int d\omega\ f(E-\omega)g(\omega)$.

For $J\gtrsim10\ \Gamma$, doing the approximations shown in Fig.~\ref{fig:SM1}, we can write:
\begin{equation}
    \bra{\beta}\hat G(\omega)\ket{\alpha}\approx
    \frac{1}{2}\left[\bra{+}\hat G(\omega)\ket{+}\pm\bra{-}\hat G(\omega)\ket{-}\right]\,,
\end{equation}
with $+$ if $\alpha=\beta$ and $-$ if $\alpha\neq\beta$. In this regime, when the dynamics are around $E\sim\omega_++\omega_-=2\omega_0$, as it is our case, the linear two-excitation Green's function after the convolution reads:
$\bra{\beta\beta}\hat G_\ell(\omega)\ket{\alpha\alpha}\approx\pm\frac{1}{2}(E-2\omega_0+i\Gamma)^{-1}$
In a similar fashion, for $J\gtrsim10\ \Gamma$, when the dynamics are around $E\sim2\omega_0$, we find:
\begin{align}
    \nonumber
    \bra{\beta\beta}&\hat G_\ell(E)\ket{a;p}\approx\pm\frac{g(p)}{2\sqrt{2}}\frac{1}{E-2\omega_0+i\Gamma}\\
    &\left[\frac{1}{E-\omega_p-\omega_++i\frac{\Gamma}{2}}+\frac{1}{E-\omega_p-\omega_-+i\frac{\Gamma}{2}}\right]\ ,
\end{align}
now with $+$ if $\beta=a$ and $-$ if $\beta=b$. Using the obtained results on Eq. \eqref{eq:B1}, we arrive to the following scattering amplitude, valid assuming only $J\gtrsim10\ \Gamma$ and $E\sim2\omega_0$:
\begin{align}
    \nonumber
    \bra{k'q'}\hat S\ket{kq}\approx&\bra{k'}\hat S\ket{k}\bra{q'}\hat S\ket{q}+\bra{q'}\hat S\ket{k}\bra{k'}\hat S\ket{q}\\
    \nonumber
    &+i\frac{\pi}{4} \ \delta(E-E')\ \frac{g^*(k')g^*(q')g(k)g(q)}{h_+(k')h_-(q')h_+(k)h_-(q)}\\
    &
    \times U\frac{E-2\omega_0+i\Gamma}{E-2\omega_0+\frac{U}{2}+i\Gamma}\ \ \ +\ \ \ 
    \begin{array}{r}
         q\leftrightarrow k \\
        q'\leftrightarrow k'
    \end{array}\ ,
\end{align}
where $h_\pm(p)=\omega_p-\omega_\pm+i\frac{\Gamma}{2}$, and $\begin{array}{cc}
     q\leftrightarrow k\\
    q'\leftrightarrow k'
\end{array}  $ represent summing all permutations. The first terms account for the linear scattering component. Interestingly, the non-linear scattering depends only on the total non-linearity $U=U_a+U_b$, and thus one of the two transmons can be replaced by a resonator and obtain the same scattering properties.
The scattering matrix shows that the inelastic scattering can only shift the individual frequencies of the photons by $|\omega-\omega'|\lesssim \Gamma$. Thus, if the right- and left-propagating input photons are of frequencies $\omega_R\sim\omega_+$ and $\omega_L\sim\omega_-$, within a bandwidth $\sigma\ll\Gamma$, the two photon scattering will be described, in the limit $\Gamma\tau\ll 1$ by:
\begin{align}
    \nonumber
   & \bra{\omega'_R\omega'_L}\hat S\ket{\omega_R\omega_L}\approx\bra{\omega'_R}\hat S\ket{\omega_R}\bra{\omega'_L}\hat S\ket{\omega_L}\\
    \nonumber
    &+i\frac{\Gamma^2}{4\pi} U\frac{E-2\omega_0+i\Gamma}{E-2\omega_0+\frac{U}{2}+i\Gamma}\ \frac{\delta(E-E')}{h_+(\omega'_R)h_-(\omega'_L)h_+(\omega_R)h_-(\omega_L)} \,,
\end{align}
where $h_\pm(\omega)=\omega-\omega_\pm+i\frac{\Gamma}{2}$, without any other process scattering into the $RR$ or $LL$ manifold, due to chirality of the couplings, see Eq.~\eqref{eq:freqdepchir}.

\end{document}